\documentclass[12pt,a4paper]{article}

\usepackage{latexsym}
\usepackage{graphicx}
\newcommand\sect[1]{\section{#1}\setcounter{equation}0}
\newcommand\ds{\displaystyle}
\newcommand\no{\nonumber\\}
\newcommand\eqnb{\begin{eqnarray}}
\newcommand\eqne{\end{eqnarray}}

\usepackage{cite}

\begin{document}

\center{{\huge The membrane as a perturbation\\  around string-like configurations}\vspace{15mm}\\
Jonas Bj\" ornsson \footnote{jonas.bjornsson@kau.se} and Stephen Hwang 
\footnote{stephen.hwang@kau.se}\\Department of Physics\\Karlstad University
\\SE-651 88 Karlstad, Sweden}\vspace{15mm}\\
\abstract{The bosonic membrane in a partial gauge, where one space dimension is eliminated,
is formulated as a perturbation theory 
around an exact free string-like solution. This perturbative regime corresponds to a situation where one
of the world-volume space-like dimensions is much greater than the other, so that the membrane has the form
of a narrow band or large hoop with string excitations being transverse to the widest dimension. 
The perturbative equations of motion are studied and solved to first order. Furthermore, it is shown 
for the open or semi-open cases 
and to any order in
perturbation theory, that one may find canonical transformations that will transform 
the membrane Hamiltonian into a free string-like
Hamiltonian and a boundary Hamiltonian. Thus the membrane dynamics in 
our perturbation scheme is essentially captured by an interacting
boundary theory defined on a two-dimensional world-sheet. A possible implication of this to M-theory is discussed.}
\newpage
\sect{Introduction}
The relativistic membrane is known to be a complicated theory even at the classical level. 
It is highly non-linear and there are only a very limited number of exact solutions known.
At the quantum level still less is known. One of the more promising approaches to the problem is the 
approximation of the membrane as a matrix model \cite{goldstone, hoppe, witt-hoppe-nicolai}. In
this approximation one may establish classically that the supermembrane in the light-cone gauge 
is the large N limit of the maximally supersymmetric $SU(N)$ matrix model. At the quantum level 
such a theory may only be consistent in $D=11$. This can be seen from the fact that, by double 
dimensional reduction \cite{duff-his}, the supermembrane contains the superstring, which is only 
consistent in
ten dimensions. A similar argument tells us that the bosonic membrane is expected to have a 
critical dimension of $27$.

The supermembrane is believed be related to the conjectured M-theory. Indeed the most fruitful definition 
hitherto of M-theory is precisely the 
large $N$ limit of the above mentioned $SU(N)$ matrix theory, a relation first conjectured in 
\cite{BFSS}. This
definition, therefore, essentially identifies M-theory and the $D=11$ supermembrane. 
Of course the discretization approach depends 
crucially on establishing that the continuum limit exists and is well-defined.

It would be useful to be able to analyze the supermembrane or M-theory directly in the continuum.
The standard approaches, so far, have been either
double-dimensional reduction, yielding a superstring theory, or by taking the field theory limit, in which
$D=11$ 
supergravity emerges. The former of these approaches may be used as a definition of M-theory 
(for early examples see eg. \cite{witten, schwarz}). None of these approximations really deal with the full supermembrane degrees of 
freedom. In this work we will propose an approximation scheme which does deal with the full 
world-volume dynamics. This scheme will formulate
the membrane as a perturbation theory around a known solution. For simplicity we will only deal with 
the bosonic membrane, and for reasons that will become clear in the following, only the open or 
semi-open cases will be of interest to us.

Our perturbation theory starts from fact that the membrane Hamiltonian, in a certain partial gauge, where 
reparametrizations in one space parameter 
are fixed and one space dimension is eliminated, may be put into
a form
\begin{equation}
H=H_0+gH_1.
\end{equation}
Here $H_0$ is the unperturbed Hamiltonian, which is of the same form as the string Hamiltonian, but with
the difference that the coordinates $X^\mu(\xi^a)$, $\mu=0,1,\ldots, D-2$, live on the world-volume 
i.e. depend on three parameters $\xi^a$, $a=0,1,2$. $H_1$ is the perturbation controlled by the parameter $g$.
As the unperturbed Hamiltonian corresponds to a solvable theory, one may use standard perturbation
theory to find the solution to any order. 

The perturbative expansion is thus 
around string-like solutions, where the membrane has the form 
of a narrow band or large hoop with string excitations being transverse to the 
largest dimension. Geometrically, this is not the natural 
string-like setting
as this would correspond to the opposite situation, where the string excitations are along the
largest dimension.
The limit $g\rightarrow 0$ corresponds to a string of infinite width/circumference.
Equivalently, this corresponds to a limit where the membrane tension is much smaller 
than the string tension associated with the string-like excitations.

Our approach is related to 
a double-dimensional reduction, since our gauge choice reduces space-time by one space dimension. 
It may, therefore,  be regarded as a perturbation around this dimensional reduction, turning on the 
dependence on the third world-volume parameter. However, this is 
not the standard double-dimensional reduction, as the latter corresponds to compactifying one dimension and 
taking the radius to zero. Our case is related to the the dual situation where the radius is very large.
In \cite{ulf, russo, sekino-yoneya} perturbative calculations around the 
double-dimensionally reduced membrane in the light-cone gauge was performed, which in spirit is 
somewhat related to our approach.

Our main focus will, however, not be to solve the perturbative equations of motion in the most 
straightforward way. Rather, we will deal with the equations by means of the Hamilton-Jacobi method, 
using canonical tranformations. We will establish the rather surprising result that the Hamiltonian of 
the partially gauge-fixed bosonic 
membrane
is, to any order in perturbation theory, canonically equivalent to the unperturbed Hamiltonian, i.e.
to the string-like Hamiltonian, complemented by a boundary Hamiltonian. Thus, in this 
perturbation scheme the membrane dynamics decomposes
into to a free "wide" string and a complicated interacting theory living on the end-lines of this string.
These end-lines sweep out two-dimensional world-sheets as they evolve in time. 

The paper is organized as follows. In the next section the perturbation theory is established. In section three, 
the classical solution to the membrane equations of motion is given to first order using straightforward 
perturbation theory. Construction of the 
solution by means of canonical transformations is done in section four and,
finally, in section five a discussion of our results and possible implications are presented.
%%%%%%%%%%%%%%%%%%%%%%%%%%%%%%%%%%%%%%%%%%%%%%%%%%%%%%%%%%%%%%%%%%%%%%%%%%%%%%%%

\sect{The basic formalism}
Let us start from the following Dirac action 
\cite{Dirac} for the membrane
\begin{equation}
S=-T_{membrane}\int d^3\xi\left[-\det\left(\partial_a X_U\partial_b X^U\right)\right]^{1/2},
\label{action}
\end{equation}
where we use the 'mostly plus' convention on the metric and $U=0,\dots,D-1$, $a=0,1,2$. 
$\xi^a$ parametrizise the worldvolume with $\xi^0=\tau$ being the time component. $T_{membrane}$ is the 
membrane tension. In passing to the 
Hamiltonian one finds the following first class constraints
\begin{eqnarray}
\phi_1&=&\mathcal P \partial_1 X \approx 0\no
\phi_2&=&\frac{1}{2}\left\{\mathcal P^2+T_{membrane}^2\left[\left(\partial_1 X\right)^2
\left(\partial_2 X\right)^2-\left(\partial_1 X\partial_2 X\right)^2\right]\right\} \approx 0\no
\phi_3&=&\mathcal P \partial_2 X \approx 0,
\end{eqnarray}
where $\mathcal P_U$ is the canonical momentum conjugate to $\dot X^U$.
Let us fix a partial gauge,
\begin{equation}
\chi\equiv X^{D-1}-\frac{T_{string}}{T_{membrane}} \xi^2 \approx 0,
\label{gaugefix}
\end{equation}
where $T_{string}$ is a constant which can be identified with a string-like tension along 
the $\xi^1$ direction. This gauge fixing may be used 
together with the constraint $\phi_3\approx 0$ to eliminate $X^{D-1}$ and 
its conjugate momentum.
Then one can write the remaining constraints as
\begin{eqnarray}
\phi_1&=&\mathcal P \partial_1 X\approx 0\no
\phi_2&=&\frac{1}{2}\left\{\mathcal P^2+T_{string}^2\left(\partial_1 X\right)^2\right.\no
&+&
\left.
T_{membrane}^2\left[
\left(\partial_1 X\right)^2\left(\partial_2 X\right)^2-\left(\partial_1 X \partial_2 X\right)^2+{1\over T_{string}^2}\left(\mathcal P 
\partial_2 X\right)^2\right]\right\} \approx 0,\no
\end{eqnarray}
where  from now on the scalar products are in $D-1$ dimensions, 
$\mathcal P \partial_1 X\equiv \mathcal P_\mu \partial_1 X^\mu$, $\mu=0,\dots,D-2$ etc. 
Let us fix $T_{string}=1$ and take $T_{membrane}\ll T_{string}=1$. Introduce a new parameter
$g\equiv\left(T_{membrane}\right)^{2}\ll 1$. By writing the 
membrane tension $T_{membrane}=T_{string} (L_2)^{-1}$, where $L_2$ gives the size in the $\xi^2$ direction,
we see that $L_2\gg 1$. Our constraints now read
\begin{eqnarray}
\phi_1&=&\mathcal P \partial_1 X\approx 0\no
\phi_2&=&\frac{1}{2}\left\{\mathcal P^2+\left(\partial_1 X\right)^2\right.\no
&+&
\left.
g\left[
\left(\partial_1 X\right)^2\left(\partial_2 X\right)^2+\left(\mathcal P 
\partial_2 X\right)^2-\left(\partial_1 X \partial_2 X\right)^2\right]\right\} \approx 0.
\label{constraints}
\end{eqnarray}
We then see  that $g$ may be 
used to define a perturbation theory. In the limit $g=0$ the constraints $\phi_1\approx 0$ and 
$\phi_2\approx 0$ reduce to the 
conventional string constraints, with the difference that $X^\mu$ and $\mathcal P^\mu$ depend on 
an additional
parameter $\xi^2$. 
By using the fundamental Poisson bracket,
\begin{equation}
\left\{X^\mu(\xi),\mathcal P^\nu(\xi')\right\}=\eta^{\mu\nu}\delta^2\left(\xi-\xi'\right),
\end{equation}
one can determine that the constraints satisfy a closed algebra,
\begin{eqnarray}
\left\{\phi_1(\xi),\phi_1(\xi')\right\}&=&\left(\phi_1(\xi)+\phi_1(\xi')\right)\partial_1\delta^2(\xi-\xi')\no
\left\{\phi_1(\xi),\phi_2(\xi')\right\}&=&\left(\phi_2(\xi)+\phi_2(\xi')\right)\partial_1\delta^2(\xi-\xi')
+g\mathcal P\partial_2 X\phi_1\left(\xi'\right)\partial_2\delta^2(\xi-\xi')\no
\left\{\phi_2(\xi),\phi_2(\xi')\right\}&=&\left(\phi_1(\xi)+\phi_1(\xi')\right)\partial_1\delta^2(\xi-\xi')\no
&+&
g\left\{\left[\left(\partial_2 X\right)^2\phi_1(\xi)+\left(\partial_2 X\right)^2\phi_1(\xi')\right]
\partial_1\delta^2(\xi-\xi')\right.\no
&-&
\left[\partial_1 X\partial_2 X\phi_1(\xi)+
\partial_1 X\partial_2 X\phi_1(\xi')\right]\partial_2\delta^2(\xi-\xi')\no
&+&
\left.
2\left[\mathcal P \partial_2 X\phi_2(\xi)+\mathcal P \partial_2 X\phi_2(\xi')
\right]\partial_2\delta^2(\xi-\xi')\right\}.
\label{constraintalgebra} 
\end{eqnarray}
We define a Hamiltonian, $H$, 
by the second constraint and separate it into two parts,
\eqnb
H_0&=&\frac{1}{2}\int d^2\xi\left[\mathcal P^2+\left(\partial_1 X\right)^2\right],
\label{unperturbed}
\\
H_1&=&\frac{1}{2}\int d^2\xi\left[\left(\partial_1 X\right)^2\left(\partial_2 X\right)^2+\left(\mathcal 
P\partial_2 X\right)^2-\left(\partial_1 X\partial_2 X\right)^2\right],
\label{perturbed}
\eqne
where $H=H_0+gH_1$. Thus $gH_1$ may be treated as a small perturbation. The gauge fixing, defined in 
equation (\ref{gaugefix}), inserted in $\phi_3=0$ yields
\eqnb
\mathcal P^{D-1}&=&-{\sqrt g} \mathcal P_\mu\partial_2 X^\mu,
\eqne
so that 
$\mathcal P^{D-1}\rightarrow 0$ when $g\rightarrow 0$. Notice also that by (\ref{gaugefix}) 
$\partial_2X^{D-1}\rightarrow \infty$ in the same limit.

Let us for completeness show that the perturbation scheme also holds for the 
BRST extended formalism. We introduce 
two ghosts $c^i$ and two anti-ghosts $b_i$ with the Poisson bracket
\begin{equation}
\left\{c^i(\xi),b_j(\xi')\right\}_+=\delta^i_j\delta^2(\xi-\xi').
\end{equation}
One may define a classical BRST charge in the standard way,
\begin{eqnarray}
Q_{BRST}&=&\int d^2\xi\left\{\phi_1c^1+\phi_2c^2+\partial_1 c^1c^1b_1+\partial_1c^2c^2b_1+
\partial_1c^2c^1b_2+\partial_1c^1c^2b_2
\right.
\no
&+&
g\left[\mathcal P\partial_2 X\partial_2c^1c^2b_1-
\partial_1 X\partial_2 X\partial_2c^2c^2b_1+\left(\partial_2 X\right)^2\partial_1c^2c^2b_1
\right.
\no
&+&
\left.\left.
2\mathcal P \partial_2 X\partial_2c^2c^2b_2+2\partial_2 c^1\partial_2 c^2c^2b_1b_2\right]\right\}.
\label{BRSTcharge}
\end{eqnarray}
From this charge one can define a BRST invariant Hamiltonian,
\begin{eqnarray}
H&=&\int d^2\xi\left\{Q_{BRST},b_2\right\}\no
&=&\int d^2\xi\left\{\phi_2-\partial_1c^2 b_1-\partial_1c^1 b_2+g\left[\partial_1 X\partial_2 X
\partial_2 c^2b_1-\mathcal P\partial_2 X\partial_2c^1b_1\right.\right.\no
&-&
\left.\left.
\left(\partial_2 X\right)^2
\partial_1 c^2b_1-2\mathcal P\partial_2 X\partial_2 c^2 b_2+
2\partial_2c^1\partial_2c^2b_1b_2\right]\right\}.
\label{hamiltonianBRST}
\end{eqnarray}
We see that the part of the Hamiltonian which contains ghosts breaks into a sum of an 
unperturbed and perturbed part,
just as the non-ghost Hamiltonian did.
%%%%%%%%%%%%%%%%%%%%%%%%%%%%%%%%%%%%%%%%%%%%%%%%%%%%%%%%%%%%%%%%%%%%%%%%%%%%%%%%%%%%%

\sect{Straightforward perturbation theory}
Let us study the equations of motion for the Hamiltonian  defined by (\ref{unperturbed}) 
and (\ref{perturbed}),
\begin{eqnarray}
\dot X^\mu &=& \mathcal P^\mu+g\partial_2 X^\mu(\mathcal P \partial_2 X),
\label{eom X}\\
\dot P^\mu &=& \partial^2_1 X^\mu +g\left\{\partial_1\left[\partial_1 X^\mu
\left(\partial_2 X\right)^2\right]+\partial_2\left[\partial_2 X^\mu\left(\partial_1 X\right)^2
\right]+\partial_2\left(\mathcal P^\mu (\mathcal P\partial_2 X)\right)\right.\no
&-&
\left.
\partial_1\left(\partial_2 X^\mu(\partial_1 X \partial_2 X)\right)-
\partial_2\left(\partial_1 X^\mu(\partial_1 X \partial_2 X)\right)\right\}.
\label{eom P}
\end{eqnarray}
To solve these equations one can make an expansion of the fields in terms of the perturbation 
parameter,
\eqnb
X^\mu=\sum_{m=0}^\infty g^m X_{(m)}^\mu\no
\mathcal P^\mu=\sum_{m=0}^\infty g^m \mathcal P_{(m)}^\mu.
\eqne
Inserting this into (\ref{eom X}), (\ref{eom P}) and separating the equation order by order yields 
to zeroth order the ordinary string equations of motion, 
\eqnb
\dot X_{(0)}^\mu&=&\mathcal P_{(0)}^\mu
\label{eom zeroth X}\\
\dot \mathcal P_{(0)}^\mu&=&\partial_1^2 X_{(0)}^\mu.
\label{eom zeroth P}
\eqne
with the solution,
\begin{equation}
X_{(0)}^\mu=X_R^\mu(\xi^0-\xi^1,\xi^2)+X^\mu_L(\xi^0+\xi^1,\xi^2).
\label{solution zeroth order}
\end{equation}
At first order the equations of motion are
\begin{eqnarray}
\dot X^\mu_{(1)}&=&\mathcal P_{(1)}^\mu+\partial_2 X_{(0)}^\mu (\mathcal P_{(0)}\partial_2 X_{(0)})
\label{1 per X}
\\
\dot \mathcal P^\mu_{(1)}&=&\partial^2_1 X_{(1)}^\mu +\left\{\partial_1\left[\partial_1 X^\mu_{(0)}
\left(\partial_2 X_{(0)}\right)^2\right]+\partial_2\left[\partial_2 X^\mu_{(0)}\left(\partial_1 X_{(0)}\right)^2
\right]\right.\no
&+&
\partial_2\left(\mathcal P_{(0)}^\mu(\mathcal P_{(0)}\partial_2 X_{(0)})\right)-
\partial_1\left(\partial_2 X^\mu_{(0)}(\partial_1 X_{(0)} \partial_2 X_{(0)})\right)\no
&-&
\left.
\partial_2\left(\partial_1 X^\mu_{(0)}(\partial_1 X_{(0)} \partial_2 X_{(0)})\right)\right\}. 
\end{eqnarray}
Eliminating $\mathcal P^\mu$ from this equation yields
\begin{eqnarray}
\Box X_{(1)}^\mu&=&-\left\{\partial_0\left[\partial_2X^\mu_{(0)}(\dot X_{(0)}
\partial_2 X_{0})\right]+\partial_1\left[\partial_1 X^\mu_{(0)}\left(\partial_2 X_{(0)}\right)^2\right]
\right.\no
&+&\partial_2\left[\partial_2 X^\mu_{(0)}\left(\partial_1 X_{(0)}\right)^2\right]
+\partial_2\left[\dot X^\mu_{(0)} (\dot X_{(0)}\partial_2 X_{(0)})\right]\no
&-&
\left.\partial_1\left[\partial_2 X^\mu_{(0)} (\partial_1 X_{(0)} \partial_2 X_{(0)})\right]
-\partial_2\left[\partial_1 X^\mu_{(0)} (\partial_1 X_{(0)} \partial_2 X_{(0)})\right]\right\},
\end{eqnarray}
where $\Box=-\partial^2_0+\partial^2_1$. To solve these equations one can introduce a Greens function,
$\Box G_S\left(\xi,\xi'\right)=\delta^2\left(\xi-\xi'\right)$, where the explicit form of 
$G_S\left(\xi,\xi'\right)$ depends on the boundary conditions. One will get the solution by 
an integration,
\begin{eqnarray}
X^\mu&=&\tilde X^\mu_R(\xi^0-\xi^1,\xi^2)+\tilde X^\mu_L(\xi^0+\xi^1,\xi^2)\no
&-&
\int d^2\xi' G_S(\xi,\xi')
\left\{\partial_0\left(\partial_2X^\mu_{(0)}(\dot X_{(0)}\partial_2 X)\right)
+\partial_1\left[\partial_1 X^\mu_{(0)}\left(\partial_2 X_{(0)}\right)^2\right]
\right.
\no
&+&
\partial_2\left[\partial_2 X^\mu_{(0)}\left(\partial_1 X_{(0)}\right)^2\right]
+\partial_2\left(\dot X^\mu_{(0)} (\dot X_{(0)}\partial_2 X_{(0)})\right)\no
&-&
\left.
\partial_1\left(\partial_2 X^\mu_{(0)} (\partial_1 X_{(0)} \partial_2 X_{(0)})\right)
-\partial_2\left(\partial_1 X^\mu_{(0)} (\partial_1 X_{(0)} \partial_2 X_{(0)})\right)\right\}
\left(\xi'\right),\no
\end{eqnarray}
where  $\tilde X^\mu_{L/R}$ is the solution 
to the homogeneous differential equation. $\mathcal P^\mu_{(1)}$ can be calculated 
using (\ref{1 per X}). One can proceed in this way to any order which will give us the 
exact solution to the equations of motion for the membrane. In the next section we will instead of 
solving the equations of motion directly show that one can use the Hamilton-Jacobi approch by 
successive canonical transformations. 
This is extensivly studied in the remaining part of this work.
%%%%%%%%%%%%%%%%%%%%%%%%%%%%%%%%%%%%%%%%%%%%%%%%%%%%%%%%%%%%%%%%%%%%%%%%%%%

\sect{Solution by canonical transformations}
In this section we will deal with the perturbative problem by means of canonical transformations.
Our approach is to find a canonical transformation, which will transform away the perturbative 
corrections, thus solving the equations of motion.

Let us begin by making a change of variables,
\begin{eqnarray}
\alpha^\mu&=&\frac{1}{\sqrt{2}}\left(\mathcal P^\mu+\partial_1 X^\mu\right)\no
\tilde\alpha^\mu&=&\frac{1}{\sqrt{2}}\left(\mathcal P^\mu-\partial_1 X^\mu\right),
\end{eqnarray}
which diagonalizes the unperturbed Hamiltonian
\eqnb
H_0&=&\frac{1}{2}\int d^2\xi\left(\alpha^2+\tilde\alpha^2\right).
\eqne
We have the Poisson bracket relations
\eqnb
\left\{\alpha^\mu(\xi),\alpha^\nu(\xi')\right\}&=&\eta^{\mu\nu}\partial_1\delta^2\left(\xi-\xi'\right)\no
\left\{\tilde\alpha^\mu(\xi),\tilde\alpha^\nu(\xi')\right\}&=&-\eta^{\mu\nu}\partial_1\delta^2\left(\xi-\xi'\right)\no
\left\{\alpha^\mu(\xi),\tilde\alpha^\nu(\xi')\right\}&=&0.
\eqne
These variables are associated with the right- and left-moving modes of the string-like configuration.
In order to express $H_1$ in terms of the new variables, we introduce a Greens function 
$K\left(\xi^1,\xi'^1\right)$ with these properties
\begin{equation}
\begin{array}{rcccl}
\ds \partial_1K\left(\xi^1,\xi'^1\right)&=&\ds -\partial'_1K\left(\xi^1,\xi'^1\right)
&=&\ds \delta\left(\xi^1-\xi'^1\right).
\label{K}
\end{array}
\end{equation}
Let us denote its operation by $\partial^{-1}_1$,
\begin{eqnarray}
\partial_1^{-1}F\left(\xi\right)&=&
\int d\xi'^1 K\left(\xi^1,\xi'^1\right)F\left(\xi^0,\xi'^1,\xi^2\right).
\eqne
It is uniquely defined up to an arbitrary $\xi^1$-independent function. From the properties of the 
Greens function we have 
$\partial_1\partial_1^{-1}F\left(\xi\right)=F\left(\xi\right)$ and 
$\partial_1^{-1}\partial_1F\left(\xi\right)=F\left(\xi\right)+f\left(\xi^0,\xi^2\right)$. 
If $F(\xi)$ is a periodic function then, 
in general, $\partial_1^{-1}F(\xi)$ is not. 
This implies that a term $\oint d\xi^1\partial_1(\ldots)$ will not necessarily be zero if the integrand contains 
terms with $\partial_1^{-1}$. This problem is basically the reason why our results do not hold for the 
closed membrane. 

Define the combination $A=\partial_2\partial_1^{-1}$ which yields 
$\partial_2X^\mu=\frac{1}{\sqrt{2}}(A\alpha-A\tilde\alpha)$ and 
\eqnb
\left\{\alpha^\mu(\xi),A\alpha^\nu(\xi')\right\}&=&\eta^{\mu\nu}\partial_2\delta^2\left(\xi-\xi'\right)\no
\left\{\tilde\alpha^\mu(\xi),A\tilde\alpha^\nu(\xi')\right\}&=&-\eta^{\mu\nu}\partial_2\delta^2\left(\xi-\xi'\right)\no
\left\{\alpha^\mu(\xi),A\tilde\alpha^\nu(\xi')\right\}&=&0\no
\left\{\tilde\alpha^\mu(\xi),A\alpha^\nu(\xi')\right\}&=&0.
\eqne
Inserting the change of variables into $H_1$ yields
\begin{eqnarray}
H_1&=&\frac{1}{8}\int d^2\xi\left[\alpha^2\left(A\alpha\right)^2+
\tilde\alpha^2\left(A\tilde\alpha\right)^2+\tilde\alpha^2\left(A\alpha\right)^2+
\alpha^2\left(A\tilde\alpha\right)^2\right.\no
&-&
\left.\left.2\alpha^2\left(A\alpha A\tilde\alpha\right)
-2{\tilde\alpha}^2\left(A\alpha A\tilde\alpha\right)-
2(\alpha{\tilde\alpha})\left(A\alpha\right)^2
-
2(\alpha{\tilde\alpha})\left(A\tilde\alpha\right)^2\right.\right.\no
&+&
4(\alpha{\tilde\alpha})\left(A\alpha A\tilde\alpha\right)
-4\left(\alpha A\alpha\right)\left(\tilde\alpha A\tilde\alpha\right)
-4\left(\alpha A\tilde\alpha\right)\left(\tilde\alpha A\alpha \right)\no
&+&
4\left.\left(\alpha A\alpha\right)\left(\tilde\alpha A\alpha\right)
+4\left(\alpha A\tilde\alpha\right)\left(\tilde\alpha A\tilde\alpha\right)
\right].
\end{eqnarray}
Let us investigate if one can make a canonical transformation 
$\left(\alpha^\mu,\tilde\alpha^\mu\right)\rightarrow\left(\alpha'^\mu,\tilde\alpha'^\mu\right)$ 
such that $H_0+gH_1\rightarrow H_0$ to first order in perturbation theory, i.e. we want a canonical 
transformation with the following property,
\eqnb
H_0\left(\alpha'^\mu,\tilde\alpha'^\mu\right)&=&H_0\left(\alpha^\mu,\tilde\alpha^\mu\right)
+g\left\{H_0\left(\alpha^\mu,\tilde\alpha^\mu\right),G_1\right\}\no
&=&H_0\left(\alpha^\mu,\tilde\alpha^\mu\right)+gH_1\left(\alpha^\mu,\tilde\alpha^\mu\right).
\label{canonical}
\eqne
One can, by inspection, directly find an expression for $G_1$ for the terms that do not mix $\alpha^\mu$ and 
$\tilde\alpha^\mu$. Assume that we have the following term in the perturbed Hamiltonian,
\eqnb
H_1^{part}&=&\int d^2\xi C_{\mu_1\dots\mu_n}\alpha^{\mu_1}\cdot\dots\cdot\alpha^{\mu_m}A\alpha^{\mu_{m+1}}
\cdot\dots\cdot A\alpha^{\mu_n},
\label{H 1 part}
\eqne
where $C_{\mu_1\dots\mu_n}$ is a constant tensor. By inspection one finds that
\eqnb
\tilde G^{part}_1&=&\int d^2\xi C_{\mu_1\dots\mu_n}\alpha^{\mu_1}\cdot\dots\cdot\alpha^{\mu_m}
A\alpha^{\mu_{m+1}}\cdot\dots\cdot A\alpha^{\mu_n}\left(\xi\right)\xi^1,
\label{G simple}
\eqne
solves equation (\ref{canonical}) up to boundary terms. This solution can be 
written as (with a suiteble choice of integration constant)
\eqnb
\tilde G^{part}_1&=&\int d^2\xi C_{\mu_1\dots\mu_n}\alpha^{\mu_1}\cdot\dots\cdot\alpha^{\mu_m}
A\alpha^{\mu_{m+1}}\cdot\dots\cdot A\alpha^{\mu_n}\left(\xi\right)\no
&&
\cdot\int d\xi'^1 K_A\left(\xi^1,\xi'^1\right),
\label{tilde G alpha}
\eqne
where $K_A(\xi^1,\xi'^1)$ is defined by
\eqnb
K_A\left(\xi^1,\xi'^1\right)&=&\frac{1}{2}
\left[K\left(\xi^1,\xi'^1\right)-K\left(\xi'^1,\xi^1\right)\right].
\label{K_A}
\eqne
This follows from the explicit form of the Greens function. 
There also exists another solution
\eqnb
G_1^{part}&=&-\int d^2\xi \int d\xi'^1 K_A\left(\xi^1,\xi'^1\right)C_{\mu_1\dots\mu_n}
\alpha^{\mu_1}\cdot\dots\cdot\alpha^{\mu_m}(\xi^0,\xi'^1,\xi^2)\no
&&
A\alpha^{\mu_{m+1}}\cdot\dots\cdot A\alpha^{\mu_n}(\xi^0,\xi'^1,\xi^2).
\label{G alpha}
\eqne
In the same manner a term of the form 
$H_1^{part}\left(\xi\right)=\int d^2\xi C_{\mu_1\dots\mu_n}\tilde\alpha^{\mu_1}\cdot\dots\cdot\tilde\alpha^{\mu_m}
\tilde  A\alpha^{\mu_{m+1}}\cdot\dots\cdot \tilde A\alpha^{\mu_n}$ is solved by
\eqnb
G^{part}_1&=&-\int d^2\xi C_{\mu_1\dots\mu_n}\tilde\alpha^{\mu_1}\cdot\dots\cdot\tilde\alpha^{\mu_m}
A\tilde\alpha^{\mu_{m+1}}\cdot\dots\cdot A\tilde\alpha^{\mu_n}\left(\xi\right)\no
&&\cdot\int d\xi'^1 K_A\left(\xi^1,\xi'^1\right)
\label{G tilde alpha}
\eqne
or
\eqnb
\tilde G_1^{part}&=&\int d^2\xi \int d\xi'^1 K_A\left(\xi^1,\xi'^1\right)C_{\mu_1\dots\mu_n}
\tilde\alpha^{\mu_1}\cdot\dots\cdot\tilde\alpha^{\mu_m}(\xi^0,\xi'^1,\xi^2)\no
&&\tilde A\alpha^{\mu_{m+1}}\cdot\dots\cdot\tilde A\alpha^{\mu_n}(\xi^0,\xi'^1,\xi^2),
\label{tilde G tilde alpha}
\eqne
where (\ref{G tilde alpha}) can be written as (up to a suitable integration constant)
\eqnb
G^{part}_1&=&-\int d^2\xi C_{\mu_1\dots\mu_n}\tilde\alpha^{\mu_1}\cdot\dots\cdot\tilde\alpha^{\mu_m}
A\tilde\alpha^{\mu_{m+1}}\cdot\dots\cdot A\tilde\alpha^{\mu_n}\left(\xi\right)\xi^1.\nonumber
\eqne
For the parts that mix different kinds of modes the situation is more complicated. Consider a 
general term of the type
\eqnb
H_1^{part}&=&\int d^2\xi C_{\mu_1\dots\mu_{q}}\alpha^{\mu_1}
\cdot\dots\cdot\alpha^{\mu_m}A\alpha^{\mu_{m+1}}\cdot\dots\cdot A\alpha^{\mu_n}\no
&&\tilde\alpha^{\mu_{n+1}}\cdot\dots\cdot\tilde\alpha^{\mu_p}A\tilde\alpha^{\mu_{p+1}}\cdot\dots\cdot
A\tilde\alpha^{\mu_q}.
\label{H^part_1}
\eqne
Let us make the following ansatz for the canonical generator 
\eqnb
G_1^{part}&=&-\frac{1}{2}\int d^2\xi C_{\mu_1\dots\mu_{q}}\int d\xi'^1 K_A\left(\xi^1,\xi'^1\right)
\alpha^{\mu_1}\cdot\dots\cdot\alpha^{\mu_m}(\xi^0,\xi'^1,\xi^2)\no
&&
A\alpha^{\mu_{m+1}}\cdot\dots\cdot A\alpha^{\mu_n}(\xi^0,\xi'^1,\xi^2)\no
&&
\tilde\alpha^{\mu_{n+1}}\cdot\dots\cdot\tilde\alpha^{\mu_p}A\tilde\alpha^{\mu_{p+1}}\cdot\dots\cdot
A\tilde\alpha^{\mu_q}\left(\xi\right),
\eqne
where $m\leq n\leq p\leq q$, $0 < n$ and $n<q$. Its Poisson bracket with $H_0$ is
\eqnb
\left\{H_0,G_1^{part}\right\}&=&-\frac{1}{2}\int d^2\xi \int d^2\xi' C_{\mu_1\dots\mu_q}
\left\{\int d\xi''^1 K_A\left(\xi'^1,\xi''^1\right)\right.\no
&&
\left[\sum_{i=1}^m\alpha^{\mu_i}(\xi)\partial_1\delta\left(\xi^1-\xi''^1\right)
\delta\left(\xi^2-\xi'^2\right)
\right.\no
&&\alpha^{\mu_1}\cdot\dots\cdot\widehat{\alpha^{\mu_i}}
\cdot\dots\cdot\alpha^{\mu_m}
A\alpha^{\mu_{m+1}}\cdot\dots\cdot A\alpha^{\mu_n}(\xi^0,\xi''^1,\xi'^2)\no
&+&
\sum_{i=m+1}^n\alpha^{\mu_i}(\xi)\delta\left(\xi^1-\xi''^1\right)
\partial_2\delta\left(\xi^2-\xi'^2\right)\no
&&
\left.
\alpha^{\mu_1}\cdot\dots\cdot\alpha^{\mu_m}A\alpha^{\mu_{m+1}}\cdot\dots\cdot 
\widehat{A\alpha^{\mu_i}}\cdot\dots\cdot A\alpha^{\mu_n}(\xi^0,\xi''^1,\xi'^2)\right]\no
&&
\tilde\alpha^{\mu_{n+1}}\cdot\dots\cdot\tilde\alpha^{\mu_p}A\tilde\alpha^{\mu_{p+1}}\cdot\dots\cdot
A\tilde\alpha^{\mu_q}(\xi^0,\xi'^1,\xi'^2)\no
&-&
\int d\xi''^1 K_A\left(\xi'^1,\xi''^1\right)\alpha^{\mu_1}
\cdot\dots\cdot\alpha^{\mu_m}(\xi^0,\xi''^1,\xi'^2)\no
&&A\alpha^{\mu_{m+1}}\cdot\dots\cdot A\alpha^{\mu_{n}}(\xi^0,\xi''^1,\xi'^2)
\no
&&
\left[
\sum_{i=n+1}^p\tilde\alpha^{\mu_i}\left(\xi\right)\partial_1\delta^2\left(\xi-\xi'\right)
\tilde\alpha^{\mu_{n+1}}
\cdot\dots\cdot\widehat{\tilde\alpha^{\mu_i}}\cdot\dots\cdot
\tilde\alpha^{\mu_p}(\xi^0,\xi'^1,\xi'^2)\right.\no
&&
A\tilde\alpha^{\mu_{p+1}}\cdot\dots\cdot
A\tilde\alpha^{\mu_q}(\xi^0,\xi'^1,\xi'^2)\no
&+&
\sum_{i=p+1}^q\tilde\alpha^{\mu_i}\left(\xi\right)\partial_2\delta^2\left(\xi-\xi'\right)
\tilde\alpha^{\mu_{n+1}}\cdot\dots\cdot\tilde\alpha^{\mu_p}(\xi^0,\xi'^1,\xi'^2)\no
&&
\left.\left.
A\tilde\alpha^{\mu_{p+1}}\cdot\dots\cdot \widehat{A\tilde\alpha^{\mu_i}}\cdot\dots\cdot
A\tilde\alpha^{\mu_q}(\xi^0,\xi'^1,\xi'^2)\right]\right\},\nonumber
\eqne
where the hat denotes an omitted term. This can be simplified using the relations in eq.(\ref{K}) and (\ref{K_A}) to read
\eqnb
\left\{H_0, G_1\right\}&=&H_1^{part}\no 
&+&
\frac{1}{2}C_{\mu_1\dots\mu_q}\left\{\int d\xi^2\left[\left(m-1\right)
\alpha^{\mu_1}\cdot\dots\cdot\alpha^{\mu_m}A\alpha^{\mu_{m+1}}
\cdot\dots\cdot A\alpha^{\mu_{n}}\left(\xi\right)\right.\right.\no
&&
\int d\xi'^1K_A(\xi^1,\xi'^1)
\tilde\alpha^{\mu_{n+1}}\cdot\dots\cdot\tilde\alpha^{\mu_p}(\xi^0,\xi'^1,\xi^2)\no
&&
A\tilde\alpha^{\mu_{p+1}}\cdot\dots\cdot A\tilde\alpha^{\mu_q}(\xi^0,\xi'^1,\xi^2)\no
&+&
\left(p-n-1\right)\tilde\alpha^{\mu_{n+1}}\cdot\dots\cdot \tilde\alpha^{\mu_p}
A\tilde\alpha^{\mu_{p+1}}\cdot\dots\cdot A\tilde\alpha^{\mu_{q}}\left(\xi\right)\no
&&
\int d\xi'^1 K_A(\xi^1,\xi'^1)
\alpha^{\mu_{1}}\cdot\dots\cdot\alpha^{\mu_m}(\xi^0,\xi'^1,\xi^2)\no
&&
\left.\left.A\alpha^{\mu_{m+1}}\cdot\dots\cdot A\alpha^{\mu_n}(\xi^0,\xi'^1,\xi^2)
\right]\right\}_{\xi^1=0}^{L_1}+\left\{\dots\right\}_{\xi^2=0}^{L_2}.
\eqne
Our ansatz yields, therefore, the correct term in the Hamiltonian. From this derivation one can also 
see that boundary terms arise. 
The solution $G_1^{part}$ is not the only one. For a Hamiltonian, as in 
eq.(\ref{H^part_1}), there are at least two linearly independent solutions,
\begin{eqnarray}
G_1^{part}&=&-\frac{1}{2}\int d^2\xi C_{\mu_1\dots\mu_{q}}\int d\xi'^1 K_A\left(\xi^1,\xi'^1\right)
\alpha^{\mu_1}\cdot\dots\cdot\alpha^{\mu_m}(\xi^0,\xi'^1,\xi^2)\no
&&
A\alpha^{\mu_{m+1}}\cdot\dots\cdot A\alpha^{\mu_n}(\xi^0,\xi'^1,\xi^2)
\tilde\alpha^{\mu_{n+1}}\cdot\dots\cdot\tilde\alpha^{\mu_p}\left(\xi\right)\no
&&
A\tilde\alpha^{\mu_{p+1}}\cdot\dots\cdot A\tilde\alpha^{\mu_q}\left(\xi\right)
\label{G mix}
\\
\tilde G_1^{part}&=&\frac{1}{2}\int d^2\xi C_{\mu_1\dots\mu_{q}}
\alpha^{\mu_1}\cdot\dots\cdot\alpha^{\mu_m}
A\alpha^{\mu_{m+1}}\cdot\dots\cdot A\alpha^{\mu_n}\left(\xi\right)\no
&&
\int d\xi'^1 K_A\left(\xi^1,\xi'^1\right)\tilde\alpha^{\mu_{n+1}}
\cdot\dots\cdot\tilde\alpha^{\mu_p}(\xi^0,\xi'^1,\xi^2)\no
&&
A\tilde\alpha^{\mu_{p+1}}\cdot\dots\cdot A\tilde\alpha^{\mu_q}(\xi^0,\xi'^1,\xi^2)
\end{eqnarray}
From here on we use the most symmetric linear combination of these two solutions
\eqnb
G_1^{sym}&=&\frac{1}{2}\left(G_1+\tilde G_1\right).
\label{symmsol}
\eqne
This will, in general, lead to simpler boundary terms. Also, we will consider an open or 
semi-open membrane. These  
cases may be solved quite generally in this approach. 

Using eqs.(\ref{tilde G alpha}), (\ref{G alpha})-(\ref{tilde G tilde alpha}) and (\ref{G mix})-(\ref{symmsol}) one finds the 
canonical generator to first order
\begin{eqnarray}
G^{sym}_1&=&-\frac{1}{32}\int d^2\xi\left[
2\partial_{1}^{-1}\left(\alpha^2\left(A\alpha\right)^2\right)
-2\alpha^2\left(A\alpha\right)^2\partial_{1}^{-1}(1)\right.\no
&+&
2\tilde\alpha^2\left(A\tilde\alpha\right)^2\partial_{1}^{-1}(1)
-2\partial_{1}^{-1}\left(\tilde\alpha^2\left(A\tilde\alpha\right)^2\right)
\no
&+&
\tilde\alpha^2\partial_1^{-1}\left(A\alpha\right)^2-
\partial_1^{-1}\tilde\alpha^2\left(A\alpha\right)^2+
\partial_1^{-1}\alpha^2\left(A\tilde\alpha\right)^2-
\alpha^2\partial_1^{-1}\left(A\tilde\alpha\right)^2
\no
&-&
2\partial_1^{-1}\left(\alpha^2A\alpha_\mu\right)A\tilde\alpha^\mu
+2\alpha^2A\alpha_\mu\partial_1^{-1}A\tilde\alpha^\mu
-2{\tilde\alpha}^2A\tilde\alpha^\mu\partial_1^{-1}A\alpha_\mu
\no
&+&
2\partial_1^{-1}\left({\tilde\alpha}^2A\tilde\alpha^\mu\right)A\alpha_\mu
-2{\tilde\alpha}^\mu\partial_1^{-1}\left(\alpha_\mu\left(A\alpha\right)^2\right)
+2\partial_1^{-1}{\tilde\alpha}^\mu\alpha_\mu\left(A\alpha\right)^2
\no
&-&
2\partial_1^{-1}\alpha_\mu{\tilde\alpha}^\mu\left(A\tilde\alpha\right)^2
+2\alpha_\mu\partial_1^{-1}\left({\tilde\alpha}^\mu A\tilde\alpha\right)^2
+4\partial_1^{-1}\left(\alpha_\mu A\alpha_\nu\right){\tilde\alpha}^\mu A\tilde\alpha^\nu
\no
&-&
4\alpha_\mu A\alpha_\nu\partial_1^{-1}\left({\tilde\alpha}^\mu A\tilde\alpha^\nu\right)
-4\partial_1^{-1}\left(\alpha_\mu A\alpha^\mu\right)\tilde\alpha_\nu A\tilde\alpha^\nu
\no
&+&
4\alpha_\mu A\alpha^\mu\partial_1^{-1}\left(\tilde\alpha_\nu A\tilde\alpha^\nu\right)
-4\partial_1^{-1}\left(\alpha_\mu A\alpha^\nu \right)A\tilde\alpha^\mu\tilde\alpha_\nu
\no
&+&
4\alpha_\mu A\alpha^\nu\partial_1^{-1}\left(A\tilde\alpha^\mu \tilde\alpha_\nu\right)
+4\partial_1^{-1}\left(\alpha_\mu A\alpha^\mu A\alpha^\nu\right)\tilde\alpha_\nu
\no
&-&
4\alpha_\mu A\alpha^\mu A\alpha^\nu\partial_1^{-1}\tilde\alpha_\nu
+
4\partial_1^{-1}\alpha_\mu A\tilde\alpha^\mu\tilde\alpha_\nu A\tilde\alpha^\nu
\no
&-&
\left.
4\alpha_\mu \partial_1^{-1}\left(A\tilde\alpha^\mu\tilde\alpha_\nu A\tilde\alpha^\nu\right)
\right],
\end{eqnarray}
where $\partial_1^{-1}$ is defined in eq.(\ref{K}), with the arbitrary function set to zero. 
Thus, what we have shown is that in place of eq.(\ref{canonical}) we find
\eqnb
H_0+g\left\{H_0,G_1\right\}&=&H_0+gH_1+gH_{B1}.
\eqne
In order to see the structure of the boundary Hamiltonian $H_{B1}$, let us collect the boundary terms 
that arise from the canonical transformation. A straightforward calculation gives the following 
explicit expression of the boundary term
\begin{eqnarray}
H_{B1}&=&\left\{\frac{1}{16}\int d\xi^2\left[
2\alpha^2\left(A\alpha\right)^2\int d\xi'^1K_A\left(\xi^1,\xi'^1\right)\right.\right.\no
&+&
2\tilde\alpha^2\left(A\tilde\alpha\right)^2\int d\xi'^1K_A\left(\xi^1,\xi'^1\right)\no
&+&
\alpha^2\int d\xi'^1K_A\left(\xi^1,\xi'^1\right)\left(A\tilde\alpha\right)^2(\xi^0,\xi'^1,\xi^2)\no
&-&
\left(A\tilde\alpha\right)^2\int d\xi'^1 K_A\left(\xi^1,\xi'^1\right)\alpha^2(\xi^0,\xi'^1,\xi^2)\no
&+&
\tilde\alpha^2\int d\xi'^1K_A\left(\xi^1,\xi'^1\right)\left(A\alpha\right)^2(\xi^0,\xi'^1,\xi^2)\no
&-&
\left(A\alpha\right)^2\int d\xi'^1 K_A\left(\xi^1,\xi'^1\right)\tilde\alpha^2(\xi^0,\xi'^1,\xi^2)\no
&-&
2\alpha^2A\alpha_\mu\int d\xi'^1 K_A\left(\xi^1,\xi'^1\right)A\tilde\alpha^\mu(\xi^0,\xi'^1,\xi^2)\no
&+&
2A\tilde\alpha_\mu\int d\xi'^1 K_A\left(\xi^1,\xi'^1\right)\alpha^2 A\alpha^\mu(\xi^0,\xi'^1,\xi^2)\no
&-&
2\tilde\alpha^2A\tilde\alpha_\mu\int d\xi'^1 K_A\left(\xi^1,\xi'^1\right)A\alpha^\mu(\xi^0,\xi'^1,\xi^2)\no
&+&
\left.\left.
2A\alpha_\mu\int d\xi'^1 K_A\left(\xi^1,\xi'^1\right)\tilde\alpha^2 A\tilde\alpha^\mu(\xi^0,\xi'^1,\xi^2)
\right]\right\}_{\xi^1=0}^{L_1}.
\label{HB}
\end{eqnarray}
Notice that the boundary Hamiltonian is non-local, since it involves several integrated Greens 
functions. Therefore, it is not a boundary in the strict sense that it depends on local functions of
the fields at the boundary. Still it is a boundary Hamiltonian in the sense 
that the corresponding Hamiltonian density is everywhere equal to its boundary value.

Let us make some comments. We have shown above $H(\alpha, \tilde\alpha)+H_{B1}(\alpha, \tilde\alpha)
=H_0(\alpha^\prime, \tilde\alpha^\prime)+O(g^2)$. This implies, however,
\eqnb 
H(\alpha, \tilde\alpha)
=H_0(\alpha^\prime, \tilde\alpha^\prime)-H_{B1}(\alpha^\prime, \tilde\alpha^\prime)+O(g^2).
\eqne
This means that the original partially gauge-fixed membrane Hamiltonian is decomposed into a string-like 
Hamiltonian and a complicated boundary Hamiltonian. This conclusion is also true to any finite
order in perturbation theory, as we will show.

Our results include the semi-open case. By choice, we can always take the boundary to have fixed 
$\xi^1$ i.e. the string-like membrane along the $\xi^1$-direction is still open and the 
two-dimensional 
surface a boundary is closed in the space-direction. Alternatively, we may mix the two 
cases having one of each type at the two boundaries. For the fully closed membrane 
the situation is different because, in this case, there do not exist any Greens functions that 
satisfies eq(\ref{K}).

Let us now show that one may extend the result to all orders in perturbation theory.
To second order the canonical transformation generated by $G_1$ is
\eqnb
f'&=&f+g\left\{f,G_1\right\}+\frac{g^2}{2}\left\{\left\{f,G_1\right\},G_1\right\},
\eqne
If we set $f=H_0$ one can see that we generate a new bulk- and 
boundary\footnote{From $G_1$ we generate new boundary terms of order $g^2$}-Hamiltonian,
\eqnb
H_0\left(\alpha'^\mu,\tilde\alpha'^\mu\right)&=&H_0\left(\alpha^\mu,\tilde\alpha^\mu\right)+
g\left[H_1\left(\alpha^\mu,\tilde\alpha^\mu\right)+H_{B1}(\alpha^\mu,\tilde\alpha^\mu)\right]\no
&+&
\frac{1}{2}g^2\left[H_2\left(\alpha^\mu,\tilde\alpha^\mu\right)+H_{B2}(\alpha^\mu,\tilde\alpha^\mu)
\right]+O(g^3).
\eqne
This can be rewritten as
\eqnb
H_0\left(\alpha'^\mu,\tilde\alpha'^\mu\right)+
g^2 H_2\left(\alpha'^\mu,\tilde\alpha'^\mu\right)=H_0\left(\alpha^\mu,\tilde\alpha^\mu\right)+
g H_1\left(\alpha^\mu,\tilde\alpha^\mu\right)+\no {\rm boundary\ terms}+ O(g^3).
\eqne
We now make a new canonical transformation generated by $G_2$ such that it compensates for the term 
$H_2$. This procedure can be continued. To $N$'th order, where $N>1$, one deduces this equation for 
the generator $G_N$
\eqnb
\left\{H_0,G_N\right\}&=&-\sum_{m,n\in \aleph,\;m \neq 1,\;nm=N}\left(-1\right)^{m}\frac{1}{m!}\;
{\mathrm ad}^m_{G_n}\left(H_0\right),
\eqne
where ${\mathrm{ad}}^m_{G_n}f\equiv\{G_n,\ldots ,\{G_n,\{G_n,f\}\ldots\}$ 
($m$~brackets). The most general bulk term of order $N$ in the Hamiltonian is of the form
\eqnb
H_N^{part}&=&\int d\xi^2 d\sigma C\left(\sigma_1,\dots,\sigma_n\right)\alpha\left(\sigma_1\right)
\cdot\dots\cdot
\alpha\left(\sigma_m\right)\tilde\alpha\left(\sigma_{m+1}\right)\cdot\dots\cdot\tilde
\alpha\left(\sigma_n\right),\no
\label{H_n^{part}}
\eqne
where we have surpressed the index structure (cf. eq.(\ref{H 1 part})) and for $\alpha$ and 
$\tilde\alpha$ we have written
\eqnb
\alpha\left(\sigma_i\right)&=&\alpha\left(\xi^0,\sigma_i,\xi^2\right).
\eqne
Let us make the ansatz
\eqnb
G_N^{part}&=&\int d\xi^2 d\sigma d\rho F\left(\sigma_1,\dots,\sigma_n,\rho_1,\dots,\rho_n\right)
C\left(\rho_1,\dots,\rho_n\right)\no
&&
\alpha\left(\sigma_1\right)\cdot\dots\cdot\alpha\left(\sigma_m\right)
\tilde\alpha\left(\sigma_{m+1}\right)
\cdot\dots\cdot\tilde\alpha\left(\sigma_n\right).
\eqne
The Poisson bracket between this ansatz and $H_0$ is,
\eqnb
\left\{H_0,G_N^{part}\right\}&=&\int d\xi^2 d\sigma d\rho \left[\sum_{i=1}^{m}\frac{\partial}{\partial\sigma_i}-
\sum_{i=m+1}^{n}\frac{\partial}{\partial\sigma_i}\right]
F\left(\sigma_1,\dots,\sigma_n,\rho_1,\dots,\rho_n\right)\no
&&
C\left(\rho_1,\dots,\rho_n\right)
\alpha\left(\sigma_1\right)
\cdot\dots\cdot\alpha\left(\sigma_m\right)
\tilde\alpha\left(\sigma_{m+1}\right)
\cdot\dots\cdot\tilde\alpha\left(\sigma_n\right).\no
\eqne
For this to be equal to (\ref{H_n^{part}}) one can see that the function $F$ has to satisfy
\eqnb
\left[\sum_{i=1}^{m}\frac{\partial}{\partial\sigma_i}-
\sum_{i=m+1}^{n}\frac{\partial}{\partial\sigma_i}\right]
F\left(\sigma_1,\dots,\sigma_n,\rho_1,\dots,\rho_n\right)&=&\prod_{i=1}^{n}\delta\left(\sigma_i-\rho_i \right).
\label{F eqn}
\eqne
The simplest way to solve this equation is to make a coordinate transformation,
\eqnb
\eta_i&=&\sum_{j=1}^{n}B_{ij}\sigma_j\no
\mu_i&=&\sum_{j=1}^{n}B_{ij}\rho_j,
\label{variable transform}
\eqne
such that the matrix $B_{ij}$ is invertible, $det\left(B_{ij}\right)=1$ and satisfies
\eqnb
\left(B^{-1}\right)_{i1}
=
\left\{
\begin{array}{cl}
1&i=1,\dots,m\\
-1&i=m+1,\dots,n.
\end{array}
\right.
\eqne
Inserting this into eq.(\ref{F eqn}) yields
\eqnb
\frac{\partial}{\partial \eta_1}
\tilde F\left(\eta_1,\dots,\eta_n,\mu_1,\dots,\mu_n\right)
&=&\prod_{i=1}^{n}\delta\left(\eta_i-\mu_i \right),
\eqne
where $\tilde F$ is related to $F$ by the variable transformation in eq.(\ref{variable transform}).
One can now use the Greens function to get the solution
\eqnb
\tilde F\left(\eta_1,\dots,\eta_n,\mu_1,\dots,\mu_n\right)
&=&
\int d\eta_1' K_A\left(\eta_1,\eta_1'\right)\delta\left(\eta_1'-\mu_1 \right)\no
&&
\cdot\prod_{i=2}^{n}\delta\left(\eta_i-\mu_i \right).
\eqne
Thus, all different kinds of terms that arise in the perturbative expansion can be solved. 
An expression for any quantity $f$ to order $N$ can be written as
\eqnb
f^{(N)}=f+
\sum_{n=1}^N 
\sum_{m=1}^{int(N/n)} (-1)^m\frac{g^{mn}}{m!}{\mathrm{ad}}^m_{G_n}f,
\eqne
where $int(N/n)$ is the integer part of $N/n$. For the Hamiltonian one can deduce that 
\eqnb H=H_0(\alpha^{(N)},\tilde\alpha^{(N)})+O(g^{N+1})+{\rm boundary\ terms}.
\eqne
Certain terms in the perturbative expansion are particularly simple to transform away. 
These are the terms that involve only one type of modes. 
For instance, if one looks at the terms involving $\alpha^\mu$ only, one can use the simple 
solution in eq.(\ref{G simple}). This generates terms of this type to second order,
\eqnb
H_2^{part}&=&\int d^2\xi D_{\mu_1\dots\mu_n}\alpha^{\mu_1}\cdot\dots\cdot\alpha^{\mu_m} 
A\alpha^{\mu_{m+1}}\cdot\dots\cdot A\alpha^{\mu_n}\left(\xi\right)\xi^1,
\eqne
where $D$ may be an operator acting on the fields. To compensate for this term one can use a 
canonical transformation generated by
\eqnb
\tilde G^{part}_2&=&\int d^2\xi D_{\mu_1\dots\mu_n}\alpha^{\mu_1}\cdot\dots\cdot\alpha^{\mu_m} 
A\alpha^{\mu_{m+1}}\cdot\dots\cdot A\alpha^{\mu_n}\left(\xi\right)\frac{\left(\xi^1\right)^2}{2}.\no
\eqne
If one proceeds to $N$'th order one finds
\eqnb
H_N^{part}&=&\int d^2\xi D_{\mu_1\dots\mu_n}\alpha^{\mu_1}\cdot\dots\cdot\alpha^{\mu_m} 
A\alpha^{\mu_{m+1}}\cdot\dots\cdot A\alpha^{\mu_n}\left(\xi\right)\xi^{N-1},
\eqne
which is solved by
\eqnb
\tilde G^{part}_N&=&\int d^2\xi D_{\mu_1\dots\mu_n}\alpha^{\mu_1}\cdot\dots\cdot\alpha^{\mu_m} 
A\alpha^{\mu_{m+1}}\cdot\dots\cdot A\alpha^{\mu_n}\left(\xi\right)\frac{\left(\xi^1\right)^N}{N}.\no
\eqne
It is interesting to note that this particularly simple solution may be applied to a string case, 
which implies that any interaction term in the Hamiltonian may be eliminated classically 
in a perturbative manner.
%%%%%%%%%%%%%%%%%%%%%%%%%%%%%%%%%%%%%%%%%%%%%%%%%%%%%%%%%%%%%%%%%%%%%%%%%%%%%%%%%%%%%%%%%%%%%

\section{Discussion}

Our treatment of the bosonic membrane, formulating it as a perturbation theory around an   
open string-like
solution, has shown to any order in perturbation theory, that the membrane equations of motions may be solved 
in the bulk of the world-volume by performing canonical transformations transforming the membrane Hamiltonian to the
free string-like Hamiltonian. At the two end-lines of the "wide" string
there remains complicated interacting theories living on  the two two-dimensional world-sheets that are traced out by 
the end-lines. These world-sheets are either
open, closed, or mixed in the space-direction, where the first possibility requires a fully open membrane. 
Of course, the bulk and boundary theories are not independent. Rather, the dynamics at the two boundaries are mediated 
by the free string oscillations of the bulk. 

It should be pointed out 
that our analysis here does not imply that the membrane theory is equivalent to a string-like theory together with a 
boundary theory, as we have only shown that, in a particular gauge, the Hamiltonians are related in this way. In order to complete the picture, 
we also need to show that the physical subspaces implied by the remaining constraints coincide. One 
may, in fact, easily 
realize that the constraints will not be canonically equivalent, not even up to boundary terms. If this would have been the
case, then the constraint algebra of the remaining constraints would have to satisfy the Virasoro algebra, which they clearly
do not when higher order terms are taken into account, as can be seen from eq.(\ref{constraintalgebra}). 
This
does not necessarily mean that the physical subspaces are inequivalent in the bulk. In order to use 
canonical transformations to
determine if the physical subspaces are equivalent, up to boundary quantities, 
one would need to extend the treatment using the membrane BRST charge. If one can show that, by extending
the canonical transformations to the ghost sector, the BRST charge of the membrane transforms into the string one and 
a boundary charge, then we are assured 
that our results here are also true for the physical solutions of the equations of motion. Notice that  
such canonical transformations mix the space-time coordinates and ghosts in a highly non-trivial way, as can 
be seen from the explicit form of the BRST charge (\ref{BRSTcharge}).
We hope to be able to report further on this issue in a forthcoming publication.

It should be remarked that from a string point of view the perturbation theory formulated here is highly non-trivial.
The canonical transformations impose corrections to the string-like modes 
$\alpha_n^\mu$ and $\tilde\alpha_n^\mu$, that have infinite net mode number, even at 
first order in perturbation theory. The  
terms that are responsible for this in the generator are the ones that mix $\alpha_n^\mu$ and $\tilde\alpha_n^\mu$. 
Consequently, the perturbative expansion 
is non-perturbative from a string point of view.

Our discussion here has been purely classical. The fact that we have formulated our perturbation theory 
around a free "wide" string means, however, that at least to zeroth order in perturbation theory we have
a quantum mechanically consistent starting point\footnote{Disregarding the usual problems of the 
bosonic string}, including a vacuum state and other physcial states. In particular, our starting point 
requires the number of space-time dimensions to be $27$. Of course, the main challenge is to see whether
this is consistent to higher orders in perturbation theory. This requires us to extend the analysis, including
the ghosts and finding first a canonical transformation which transforms the BRST charge (\ref{BRSTcharge}),
modified by boundary terms, to the string BRST charge, as was discussed above. Then, if this transformation 
may be extended to a unitary one
at the quantum level, consistency at $D=27$ is established. It might seem that the hope of showing that the 
canonical transformation extends into a unitary one is very optimistic. However, the main problem is one 
of ordering operators and this may be solved by requiring that the ordering defined by consistency of the string-like
solution, will define the ordering of the operators prior to transformation. 

We have here treated the bosonic membrane. The most interesting case is, however, the supermembrane. It remains to 
see whether our treatment extends to this case as well. Our belief is that this indeed is the case. There seems 
to be no principle difference between the two when it comes to formulating the perturbation theory.
However, it is well-known that the 
two cases differ in many respects. One of the more important aspects is that the bosonic membrane has a discrete
spectrum, whereas the supermembrane has a continous one \cite{witt-luscher-nicolai}. This has important consequences in the 
interpretation of the latter theory (see eg. the discussion in \cite{nicolai-helling}). Whether a similar 
treatment of the supermembrane will highlight these differences remains to be seen. It may turn out 
that the complexity of the boundary Hamiltonian prevents any further understanding in this respect.

Let us end this discussion with an even more speculative remark. As we have pointed out several times the 
non-trivial part of the membrane theory, in our scheme, are the boundary theories living on the two-dimensional 
world-sheets at the end-lines. It would be tempting to say that each of these latter theories
correspond to some sort of interacting string theory. For this to be true, it is necessary that the membrane 
theory, supplemented by boundary theories, are reparametrication invariant also at the boundaries.  
Taking such a fully invariant theory our perturbative scheme may imply that this theory is canonically equivalent to  
an interacting open or closed string theory at each boundary, which communicate with each other through the free
"wide" string in the bulk. 
We could go one step further and say that M-theory, in a partial gauge, may perhaps be {\it defined} in this way. 
With such a point 
of view,  M-theory dynamics would essentially
reduce to that of two coupled interacting string theories. Such a definition would make it
possible to analyze M-theory in great detail.


\begin{thebibliography}{99}
\bibitem{goldstone}J. Goldstone, unpublished
\bibitem{hoppe} J. Hoppe, {\it Quantum theory of a massless Relativistic Surface and a two-dimensional 
bound state problem}, PhD thesis MIT 1982
%\bibitem{bergshoeff-sezgin-tanii}E. Bergshoeff, E. Sezgin and Y. Tanii, Nucl. Phys. {\bf B298} (1988) 187
\bibitem{witt-hoppe-nicolai}B. de Witt, J. Hoppe and H. Nicolai, Nucl. Phys. {\bf B305} (1988) 545
\bibitem{duff-his}M.J. Duff, P.S. Howe, T. Inami and K.S. Stelle, Phys. Lett. 191 B, 70, 1987
\bibitem{BFSS}T.Banks, W Fischler, S.H. Shenker and L. Susskind, Phys. Rev {\bf D55} (1997) 5112, 
hep-th/9610043
\bibitem{witten} E. Witten, Nucl. Phys. {\bf B443} (1995) 85-126, {hep-th/9503124}
\bibitem{schwarz} J. Schwarz, Phys. Lett. {\bf B367} (1996) 97, hep-th/9510086
%\bibitem{duff-inami-pope-sezgin}M.J. Duff, T. Inami, C.N. Pope, E. Sezgin and K.S. Stelle, Nucl.\ Phys.\ {\bf B297} (1988) 515.
\bibitem{ulf}U. Lindstr\"om, Phys.\ Lett.\ {\bf B218} (1989) 315.
\bibitem{russo}J.G. Russo, Nucl.\ Phys.\ {\bf B492} (1997) 205, {hep-th/9610018}.
\bibitem{sekino-yoneya}Y. Sekino and T. Yoneya, Nucl.Phys. {\bf B619} (2001) 22, {hep-th/0108176}
\bibitem{Dirac} P. A. M. Dirac, Proc. Roy. Soc. {\bf A268} 57 (1962)
\bibitem{witt-luscher-nicolai}B. de Witt, M. L\"uscher and H. Nicolai, Nucl. Phys. {\bf B320} (1989) 135
\bibitem{nicolai-helling}H. Nicolai and R. Helling, {Supermembranes and M(atrix) theory}, hep-th/9809103
\end{thebibliography}
\end{document}